\documentclass[10pt,final,journal,letterpaper,oneside,twocolumn]{IEEEtran}
\usepackage[pdftex]{graphicx}
\usepackage[ruled]{algorithm2e}
\usepackage{algorithmic}
\usepackage[small]{caption}
\usepackage{float}
\usepackage{xspace}
\usepackage{fancyhdr}
\usepackage{amssymb,amsmath}
\usepackage{subfig}
\usepackage{epstopdf}
\usepackage{bbding}
\usepackage{cite} 
\usepackage{color}
 \usepackage[english]{babel}
\usepackage{amsthm}

\begin{document}
%
\title{Enhancing Physical Layer Security in Cognitive Radio-Enabled NTNs with Beyond Diagonal RIS}

\author{Wali Ullah Khan, \textit{Member, IEEE,} Chandan Kumar Sheemar, \textit{Member, IEEE,}\\ Eva Lagunas, \textit{Senior Member, IEEE,} Symeon Chatzinotas, \textit{Fellow, IEEE}  \thanks{Authors are with the Interdisciplinary Centre for Security, Reliability, and Trust (SnT), University of Luxembourg, 1855 Luxembourg City, Luxembourg (emails: \{waliullah.khan, eva.lagunas, chandankumar.sheemar, symeon.chatzinotas\}@uni.lu).

}}%

\markboth{IEEE Conference
}
{Shell \MakeLowercase{\textit{et al.}}: Bare Demo of IEEEtran.cls for IEEE Journals} 

\maketitle

\begin{abstract}
Beyond diagonal reconfigurable intelligent surfaces (BD-RIS) have emerged as a transformative technology for enhancing wireless communication by intelligently manipulating the propagation environment. This paper explores the potential of BD-RIS in improving cognitive radio enabled multilayer non-terrestrial networks (NTNs). It is assumed that a high-altitude platform station (HAPS) has set up the primary network, while an uncrewed aerial vehicle (UAV) establishes the secondary network in the HAPS footprint. We formulate a joint optimization problem to maximize the secrecy rate by optimizing BD-RIS phase shifts and the secondary transmitter power allocation while controlling the interference temperature from the secondary network to the primary network. To solve this problem efficiently, we decouple the original problem into two sub-problems, which are solved iteratively by relying on alternating optimization. Simulation results demonstrate the effectiveness of BD-RIS in cognitive radio-enabled multilayer NTNs to accommodate the secondary network while satisfying the constraints imposed from the primary network.
\end{abstract}


\IEEEpeerreviewmaketitle

\section{Introduction}
Integrating non-terrestrial networks (NTNs) into 6G technology delivers uninterrupted global coverage, ubiquitous connectivity, highly reliable communication, and improved network accessibility in remote or underserved regions \cite{10716670}. In contrast to conventional terrestrial setups, NTNs utilize satellites, high-altitude platform stations (HAPS), and uncrewed aerial vehicles (UAVs) to expand network reach beyond traditional boundaries \cite{9861699}. These multi-tiered architectures facilitate resilient wireless communication across varied terrains, enabling advanced applications such as the Internet of Things (IoT), connected vehicle systems, and disaster response mechanisms \cite{10097680}.

Nevertheless, despite their significant promise, NTNs encounter numerous challenges, including dynamic spectrum sharing, substantial signal attenuation, rapidly changing channel environments, and energy inefficiencies caused by the vast distances between transmitters and receivers \cite{10715713,10684731}. Additionally, ensuring robust physical layer security remains a critical concern, as the open nature of wireless communication in NTNs makes them vulnerable to eavesdropping and other malicious attacks \cite{10003076}. This motivates the need for solutions which jointly take into account the security requirements while considering also the integration of different types of networks coexisting in the same spectrum.

To tackle these challenges, cognitive radio networks and beyond diagonal reconfigurable intelligent surfaces (BD-RIS) have emerged as pivotal technologies for enhancing spectrum efficiency and boosting network adaptability \cite{10816483, 9514409,10834443}. Cognitive radio networks function with a primary network and an underlaid secondary network, allowing the latter to opportunistically access underutilized frequency bands while ensuring controlled interference temperature towards the primary network \cite{10013700}. This dynamic spectrum access approach effectively addresses spectrum scarcity in integrated terrestrial and NTNs, enabling more flexible and efficient communication and enhanced spectrum efficiency. On the other hand, BD-RIS provides advanced beamforming capabilities for interference mitigation, which can be achieved by dynamically reconfiguring the wireless environment through interconnected reconfigurable elements, leading to a fully controllable scattering matrix \cite{9913356}. By integrating cognitive radio and BD-RIS, 6G NTNs can achieve greater spectral efficiency, enhanced energy efficiency, and more robust communication performance, even in complex and demanding environments \cite{khan2024integration}.

Existing literature on secrecy rate optimization in cognitive radio networks has primarily focused on diagonal RIS, where passive beamforming is limited to independent phase shift adjustments. For example,  \cite{9946982} formulated a joint secrecy rate and transmission rate optimization problem, leveraging RIS to minimize transmit power while ensuring physical layer security. The authors used an iterative algorithm based on second-order cone form to obtain an efficient solution and validated the proposed solution with numerical results. The study in \cite{10333515} considered robust transmission design under imperfect CSI uses alternating optimization and convex approximation methods to improve energy and spectrum efficiency. In \cite{10683052}, the secrecy performance of a RIS-assisted D2D communication system was also analyzed, addressing energy harvesting and outage probabilities in a cognitive cellular network. Additionally, \cite{10883339} jointly optimized the secure transmission of both primary and secondary users in an RIS-aided cognitive radio network, proposing alternating optimization and semi-definite relaxation techniques to tackle security threats from eavesdroppers. Furthermore, in \cite{10460375}, RIS was employed to enhance spectrum sensing accuracy and secrecy performance in cognitive radio networks, utilizing a block coordinate descent-based approach to optimize beamforming and phase shifts, effectively improving both sensing efficiency and physical layer security.

Note that the potential of BD-RIS for physical layer security has yet to be explored. The unique ability of BD-RIS to dynamically manipulate electromagnetic waves provides an opportunity to enhance security measures against eavesdropping and jamming attacks. By leveraging the increased degrees of freedom offered by BD-RIS, along with advanced signal processing techniques, it is possible to achieve substantial improvements in secure communications. 
 
In this work, we investigate the physical layer security of a cognitive radio-enabled NTN with a transmissive BD-RIS-mounted UAV serving as a secondary transmitter (ST), while an HAPS exists as a primary transmitter. We aim at maximizing the secrecy rate of the secondary network which adhering the interference constraint, linked through the interference temperature, imposed by the primary network. A joint optimization framework is proposed to maximize the secrecy rate of the secondary network by jointly optimizing the BD-RIS's phase response and the transmit power at the secondary network. However, remark that this results in a highly non-convex optimization problem which is very challenging to be solved. To overcome this hurdle, we rely on an alternating optimization, by decomposing the global optimization problem into a series of two sub-problems, each solved for one variable while keeping the remaining variables fixed. The proposed solution iterates between a low-complexity switch-based solution for optimal power allocation and Stiefel manifold
optimization for the BD-RIS phase response. Simulation results demonstrate the effectiveness of the proposed approach in significantly improving the performance of the multi-layered network in terms of secrecy rate, compared to the traditional diagonal RIS.

\emph{Paper Organization:} The rest of the paper is organized as follows. In Section \ref{sec_2}, the system model for the cognitive-radio enabled multi-layered NTN is presented and the secrecy rate maximization problem is formulated. In Section \ref{sec_3}, a novel optimization framework to enhance the physical layer security is presented. Finally, in Section \ref{sec_4} and \ref{sec_5}, simulation results and conclusions are presented. 
\begin{figure}[!t]
\centering
\includegraphics [width=.45\textwidth]{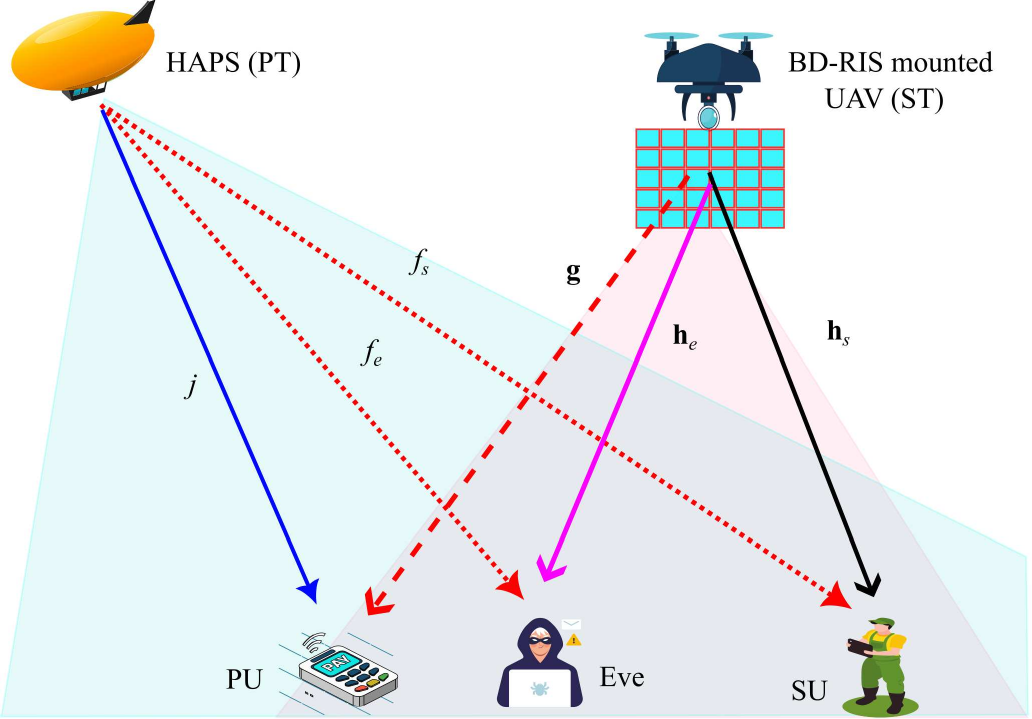}
\caption{System model.}
\label{PLSsm}
\end{figure}

\section{System Model and Problem Formulation} \label{sec_2}
We consider a cognitive radio-enabled NTN that consists of a primary network with the HAPS primary transmitter (PT) serving a primary user (PU) and a transmissive BD-RIS-mounted UAV secondary transmitter (ST) serving a secondary user (SU) as shown in Fig. \ref{PLSsm}. BD-RIS consists of \(M\) reconfigurable elements and follows a fully-connected architecture with interconnected elements. The primary and secondary networks share the same spectrum to maximize spectral efficiency. A passive Eve is present in the secondary network, attempting to intercept the signal intended for the SU. The Eve also receives interference from the primary network, which affects its ability to decode the intercepted signal. The objective is to maximize the secrecy rate of the secondary UAV network by optimizing the transmit power of the ST and the phase shift matrix of the BD-RIS, subject to interference constraints at the PU. We consider the UAV to be in a nearly stationary position while hovering \cite{9515574}. Therefore, the channel vector $\mathbf{h}_s\in \mathbb{C}^{M \times 1}$ from ST to SU can be modeled as Rician fading such as:
\begin{equation}
\mathbf{h}_s = \sqrt{\frac{\hat{h}_s}{d_s^2}}\Bigg (\sqrt{\frac{K}{K+1}} \mathbf{h}_{s,\text{LoS}} + \sqrt{\frac{1}{K+1}} \mathbf{h}_{s,\text{NLoS}}\Bigg),\label{1}
\end{equation}
where \(K\) is the Rician factor, $\hat{h}_s$ denotes the channel power gain, $d_s$ shows the distance between ST and SU, \(\mathbf{h}_{s,\text{LoS}}\) represents the deterministic line-of-sight (LoS) component, and \(\mathbf{h}_{s,\text{NLoS}}\) follows a Rayleigh distribution. According to \cite{10839492}, the LoS component \(\mathbf{h}_{s,\text{LoS}}\) can be further expressed as:
\begin{align}
&\mathbf{h}_{s,\text{LoS}}=[1,\exp^{-j\delta\sin{\theta}\cos{\varphi}},\dots,\exp^{-j\delta\sin{\theta}\cos{\varphi}(M_x-1)}]^T\nonumber\\& \otimes[1,\exp^{-j\delta\sin{\theta}\sin{\varphi}},\dots,\exp^{-j\delta\sin{\theta}\sin{\varphi}(M_y-1)}]^T
\end{align}
where \(\delta=(2\pi f_c q)/c\) with $q$ being the distance between adjacent phas shift elements.

Similarly, the channel from ST to Eve can be modeled as:
\begin{equation}
\mathbf{h}_e = \sqrt{\frac{\hat{h}_e}{d_e^2}}\Bigg (\sqrt{\frac{K}{K+1}} \mathbf{h}_{e,\text{LoS}} + \sqrt{\frac{1}{K+1}} \mathbf{h}_{e,\text{NLoS}}\Bigg)\in \mathbb{C}^{M \times 1}.
\end{equation}
where the LoS component \(\mathbf{h}_{e,\text{LoS}}\) can be described as:
\begin{align}
&\mathbf{h}_{e,\text{LoS}}=[1,\exp^{-j\delta\sin{\theta}\cos{\varphi}},\dots,\exp^{-j\delta\sin{\theta}\cos{\varphi}(M_x-1)}]^T\nonumber\\& \otimes[1,\exp^{-j\delta\sin{\theta}\sin{\varphi}},\dots,\exp^{-j\delta\sin{\theta}\sin{\varphi}(M_y-1)}]^T
\end{align}
Subsequently, the channel from ST to PU and PT to SU and Eve follows Rician fading and can be modeled similarly to (\ref{1}); however, the detail is omitted here due to limited space and simplicity. The received signal of SU and Eve from ST can be defined as: 
\begin{equation}
    y_s = \mathbf{h}_s \mathbf{\Phi}\sqrt{P_s} x +f_s\sqrt{Q_p}z+ n_{s},
\end{equation}
\begin{equation}
    y_e = \mathbf{h}_e \mathbf{\Phi}\sqrt{P_s} x +f_e\sqrt{Q_p}z+ n_{e},
\end{equation}
where \(x\) is the transmitted signal from the ST to SU, $z$ is the transmitted signal of PT for PU, $\boldsymbol{\Phi}\in\mathcal C^{M_x\times M_y}$ denotes the phase shift matrix of transmissive BD-RIS such that $\boldsymbol{\Phi}\boldsymbol{\Phi}^H={\bf I}_M$, where $M_x$ is the number of elements in each column and $M_y$ is the elements in each row, and \(n_{s} \sim \mathcal{CN}(0, \sigma_s^2)\) and \(n_{e} \sim \mathcal{CN}(0, \sigma_e^2)\) are the additive white Gaussian noise (AWGN). Furthermore, $P_s$ and $Q_p$ show the transmit power of ST and PT. The signal to interference plus noise ratio (SINR) for SU and Eve can be expressions are: 
\begin{align}
\gamma_s &= \frac{|\mathbf{h}_s \mathbf{\Phi}|^2 P_{s} }{\sigma_s^2+|f_s|^2Q_p}, \label{4} \\
\gamma_e &= \frac{|\mathbf{h}_e \mathbf{\Phi}|^2 P_{s} }{\sigma_e^2+|f_e|^2Q_p},
\end{align}
where $\sigma_s,\sigma_e$ represent the variance of SU and Eve. The secrecy rate of the secondary UAV network is defined as:
\begin{equation}
    R_{\text{sec}} = \max \left( 0, R_s - R_e \right),
\end{equation}
where $R_s = \log_2 \left( 1 + \gamma_s \right)$ and $R_e = \log_2 \left( 1 + \gamma_e \right)$. To ensure the service quality of the primary network, the interference power at the PU from ST is constrained by the interference temperature limit \(I_{\text{th}}\) such as: 
\begin{equation}
 |\mathbf{g} \mathbf{\Phi}|^2 P_{s} \leq I_{\text{th}}.
\end{equation}

This work seeks to maximize the secrecy rate of the secondary UAV network by optimizing the transmit power of UAV and the phase shift matrix of BD-RIS. The optimization problem for secrecy rate maximization can be formulated as:
\begin{align}
 (\mathcal{P}) \quad   \max_{P_{s}, \mathbf{\Phi}} \quad & R_s^{\text{sec}} = [\log_2 \left( 1 + \gamma_s \right) - \log_2 \left( 1 + \gamma_e \right)]^+ \label{11}\\
    \text{s.t.} \quad C_1:\ & |\mathbf{g} \mathbf{\Phi}|^2 P_{s} \leq I_{\text{th}}, \label{12}\\
    C_2:\ & 0 \leq P_{s} \leq P_{\max},\label{13} \\
   C_3:\ & \mathbf{\Phi}\mathbf{\Phi}^H = \textbf{I}_m,\label{14}
\end{align}
where (\ref{11}) aims to maximize secrecy rate, $C_1$ in (\ref{12}) ensures the PU’s QoS, $C_2$ in (\ref{13}) limits the ST’s transmit power, and $C_3$ in (\ref{14}) enforces BD-RIS phase shift constraints. The problem $(\mathcal{P})$ is non-convex due to coupling between $P_s$ and $\boldsymbol{\Phi}$, so we solve it efficiently using alternating optimization.

\section{Proposed Solution} \label{sec_3}
The problem $(\mathcal{P})$ is non-convex due to the coupling between the power allocation and the phase shift matrix, as well as the unitary constraint of BD-RIS, making it challenging to obtain a joint optimal solution. To solve the optimization problem \((\mathcal{P})\) efficiently, we adopt the alternating optimization. This approach decouples the problem into two subproblems: optimizing the transmit power \(P_s\) of the UAV and optimizing the phase shift matrix \(\mathbf{\Phi}\) of the BD-RIS.

\subsection{Power Allocation}
Given the fixed phase shift matrix at BD-RIS, the optimization problem \((\mathcal{P})\) can be simplified to the power allocation problem \(P_s\) as:
\begin{align}
 (\mathcal{P}1) \quad   \max_{P_{s}} \quad & [\log_2 \left( 1 + \gamma_s \right) - \log_2 \left( 1 + \gamma_e \right)]^+ \label{obj}\\
    \text{s.t.} \quad C_1:\ & |\mathbf{g} \mathbf{\Phi}|^2 P_{s} \leq I_{\text{th}}, \label{16}\\
    C_2:\ & 0 \leq P_{s} \leq P_{\max},\label{17} 
\end{align}
We can notice that this problem lacks a closed-form solution that meets the first-order optimality conditions, as both terms increase monotonically with respect to $P$, resulting in stationarity conditions that cannot be resolved.
Moreover, it can be observed in problem $(\mathcal{P}1)$ that the objective function \([\log_2 \left( 1 + \gamma_s \right) - \log_2 \left( 1 + \gamma_e \right)]^+\) is monotonically increasing with respect to \(P_s\) through \(\gamma_s\) and \(\gamma_e\). However, the difference \([\log_2 \left( 1 + \gamma_s \right) - \log_2 \left( 1 + \gamma_e \right)]^+\) depends on the relative strengths of the normalized channel to interference plus the noise gain \(\frac{h_s}{\sigma_s^2+|f_s|^2Q_p}\) and \(\frac{h_e}{\sigma_e^2+|f_e|^2Q_p}\), where \(h_s=|\mathbf{h}_s \mathbf{\Phi}|^2\) and \(h_e=|\mathbf{h}_e \mathbf{\Phi}|^2\). For example, if
\begin{align}
\frac{h_s}{\sigma_s^2+|f_s|^2Q_p}>\frac{h_e}{\sigma_e^2+|f_e|^2Q_p},
\end{align}
the term \(\log_2 \left( 1 + \gamma_s \right)\) will dominate and increasing the transmit power will increase the secrecy rate of the secondary UAV network. The optimal power \(P_s\) in this case can be determined by the interference constraint in (\ref{16}) and the maximum power constraint in (\ref{17}). It can be expressed as:
\begin{align}
P_s^* = \min \left( \frac{I_{\text{th}}}{|\mathbf{g} \mathbf{\Phi}|^2}, P_{\max} \right).
\end{align}
In case if 
\begin{align}
\frac{h_s}{\sigma_s^2+|f_s|^2Q_p}<\frac{h_e}{\sigma_e^2+|f_e|^2Q_p},
\end{align}
the term \(\log_2 \left( 1 + \gamma_e \right)\) become equal or higher than \(\log_2 \left( 1 + \gamma_s \right)\), which result in zero secrecy rate. In such a scenario, the optimal approach is to transmit zero power, i.e., \(P^*_s=0\). Overall, the optimal power can be expressed as:
\begin{equation}
    P_s^*=\begin{cases}
    \min \left( \frac{I_{\text{th}}}{|\mathbf{g} \mathbf{\Phi}|^2}, P_{\max} \right), \ \text{if} \ \frac{h_s}{\sigma_s^2+|f_s|^2Q_p}>\frac{h_e}{\sigma_e^2+|f_e|^2Q_p},\\
    0, \ \text{if}\ \frac{h_s}{\sigma_s^2+|f_s|^2Q_p}<\frac{h_e}{\sigma_e^2+|f_e|^2Q_p}.
    \end{cases}
\end{equation}
Note that this solution means that the optimal transmit power $P^*_s$ depends on the channel gains and the interference temperature from secondary to primary network.

\subsection{Phase Shift Matrix Design}
For a given transmit power at UAV $P_s$, the optimization problem $(\mathcal{P})$ reduces to a phase shift matrix design problem as: 
\begin{align}
 (\mathcal{P}2) \quad   \max_{\mathbf{\Phi}} \quad & R_s^{\text{sec}} =[\log_2 \left( 1 + \gamma_s \right) - \log_2 \left( 1 + \gamma_e \right)]^+ \label{obj}\\
    \text{s.t.} \quad C_1:\ & |\mathbf{g} \mathbf{\Phi}|^2 P_{s} \leq I_{\text{th}},\\
   C_3:\ & \mathbf{\Phi}\mathbf{\Phi}^H = \textbf{I}_m,\label{C3}
\end{align} 
The above problem is non-convex due to the objective function, which involves the difference of two logarithmic functions. Moreover, $\gamma_s$ and $\gamma_e$ are quadratic functions of $\mathbf{\Phi}$, making the objective function highly non-linear and non-convex. Furthermore, the unitary constraint is non-linear and non-convex, further complicating the problem. To address this, we adopt Manifold optimization which is a powerful tool for solving optimization problems with unitary constraints. In this method, $\mathbf{\Phi}$ is treated as a point on the Stiefel manifold, which is the set of all $M\times M$ unitary matrices. The Stiefel manifold $St(M,M)$ can be defined as:
\begin{align}
St(M,M)=\{\boldsymbol{\Phi}\in \mathcal{C}^{M\times M}|\mathbf{\Phi}\mathbf{\Phi}^H = \textbf{I}_m\}.\label{24}
\end{align}
Now we introduce a Lagrange multiplier $\lambda$ and define an augmented Lagrangian function as:
\begin{align}
L(\boldsymbol{\Phi},\lambda)=R^{\text{sec}}_s-\lambda(|\mathbf{g} \mathbf{\Phi}|^2 P_{s}-I_{\text{th}}).\label{new}
\end{align}
Next, we compute the Euclidean gradient of the objective function $R_s^{\text{sec}}$ with respect to $\boldsymbol{\Phi}$, which can be described as:
\begin{align}
&\nabla_{\mathbf{\Phi}}L(\Phi, \lambda) = -2\lambda P_s(\mathbf{g}^H\boldsymbol{\Phi})\mathbf{g}  + \frac{2 P_s}{\ln(2)}\times\label{25}\\& \left( \frac{\mathbf{h}_s^H A_s}{(1 + \gamma_s)(\sigma^2 + |f_s|^2 Q_p)} - \frac{\mathbf{h}_e^H A_e}{(1 + \gamma_e)(\sigma^2 + |f_e|^2 Q_p)} \right),\nonumber
\end{align}
where
\(
A_s = \mathbf{h}_s \mathbf{\Phi},\ A_e = \mathbf{h}_e \mathbf{\Phi}.
\) The detailed derivation of (\ref{25}) can be found in Appendix A.
The Euclidean gradient provides the direction of steepest ascent in the ambient Euclidean space. However, this gradient does not respect the manifold structure, so it must be projected onto the tangent space of the Stiefel manifold. The next step is to project the Euclidean gradient onto the tangent space of the Stiefel manifold at the current point $\boldsymbol{\Phi}_k$ as:
\begin{align}
&\text{grad}_{\boldsymbol{\Phi}}L=\nabla_{\boldsymbol{\Phi}}L-\boldsymbol{\Phi}_k(\nabla_{\boldsymbol{\Phi}}L)^H\boldsymbol{\Phi}_k,
\end{align}
where $k$ denotes the iteration index. If the interference threshold is violated at any iteration, project the update back onto the feasible region by adjusting the step size $I_{\text{th}}$ as:
\begin{align}
\eta_k=\min\bigg(\eta_k,\frac{I_{\text{th}}}{|\textbf{g}\Phi|^2P_s}\bigg).
\end{align}
The tangent space is a linear approximation of the manifold at $\boldsymbol{\Phi}_k$, and the projection ensures that the gradient direction respects the manifold's geometry. This connects the Euclidean gradient in (\ref{25}) to the manifold structure in (\ref{24}), ensuring that updates are consistent with the unitary constraint. We update $\boldsymbol{\Phi}$ on the manifold using a matrix exponential method as:
\begin{align}
\boldsymbol{\Phi}_{k+1}=\boldsymbol{\Phi}_k\text{Exp}(\eta_k.\text{gad}_{\boldsymbol{\Phi}}R_s^{\text{sec}}), \label{27}
\end{align}
This method updates $\boldsymbol{\Phi}$ iteratively, refining the phase shift matrix until convergence is achieved. The step size $\eta_k$ controls the movement along the geodesic path of the manifold, ensuring a stable and efficient optimization process. The process from (\ref{new})–(\ref{27}) is repeated until the algorithm converges to an optimal phase shift matrix.

\section{Numerical Results} \label{sec_4}
This section presents numerical results of the proposed framework, obtained through Monte Carlo simulations. Unless otherwise specified, the simulation parameters are set as follows: the number of BD-RIS elements varies from \( M = 8 \) to \( M = 64 \), the noise variance is \( \sigma^2 = 0.0001 \), the interference temperature limit is \( I_{\text{th}} = 10^{-5} \), the maximum transmit power of the UAV ranges from \( P_{\text{max}} = 20 \) dBm to \( P_{\text{max}} = 30 \) dBm, and the transmit power of the primary transmitter (PT) is \( Q_p = 40 \) dBm. The distances are set as follows: 800 m from PT to Eve, 1000 m from PT to the secondary user (SU), 100 m from the UAV to SU, and 110 m from the UAV to both Eve and the primary user (PU). The Rician factor is set to 5 for the channels from PT to SU and Eve, and 10 for the channels from the UAV to SU, Eve, and PU. The number of Monte Carlo trials is 10,000. We compare the proposed optimization framework with a benchmark method where the phase shift matrix of BD-RIS is randomly initialized.
\begin{figure}[!t]
\centering
\includegraphics [width=.48\textwidth]{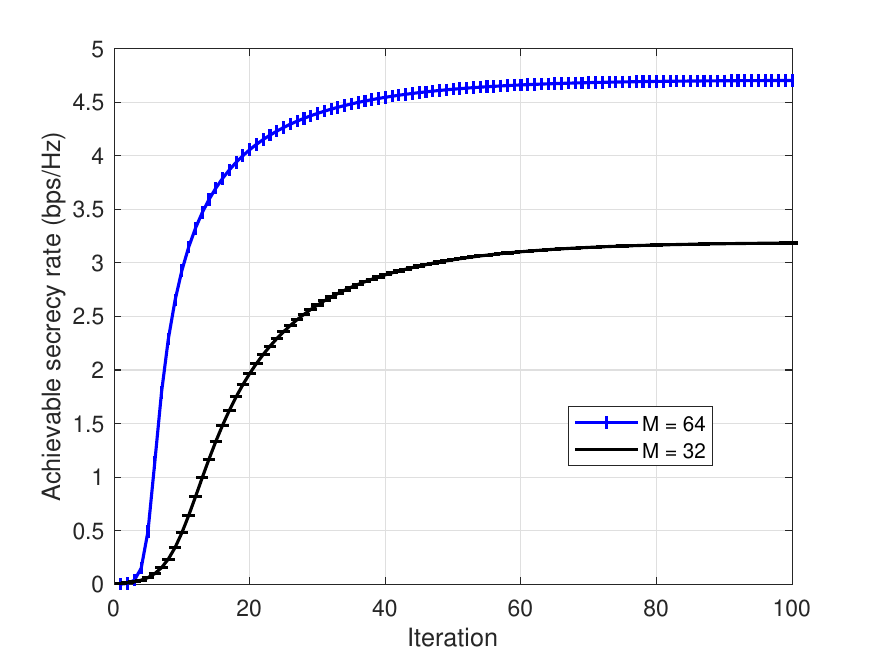}
\caption{Number of iterations versus achievable secrecy rate of the system, considering $P_{\text{max}}=20$ dBm and $I_{\text{th}}=10^{-5}$.}
\label{converg}
\end{figure}

Fig. \ref{converg} illustrates the convergence of the proposed framework in terms of the achievable secrecy rate (bps/Hz) over iterations for different numbers of BD-RIS elements, i.e., $M=32$ and $M=64$. The results indicate that as the number of iterations increases, the secrecy rate improves significantly before stabilizing. In particular, a BD-RIS with $M=64$ leads to a higher achievable secrecy rate than $M=32$M = 32, highlighting the benefit of increasing the bd-RIS size in enhancing physical layer security. This improvement stems from the additional degrees of freedom the BD-RIS provides, enabling better interference management and secure beamforming. Furthermore, the convergence behavior suggests that the proposed framework efficiently optimizes the secrecy rate within a reasonable number of iterations, making it practical for real-time implementation in cognitive radio-enabled NTNs. 
\begin{figure}[!t]
\centering
\includegraphics [width=.48\textwidth]{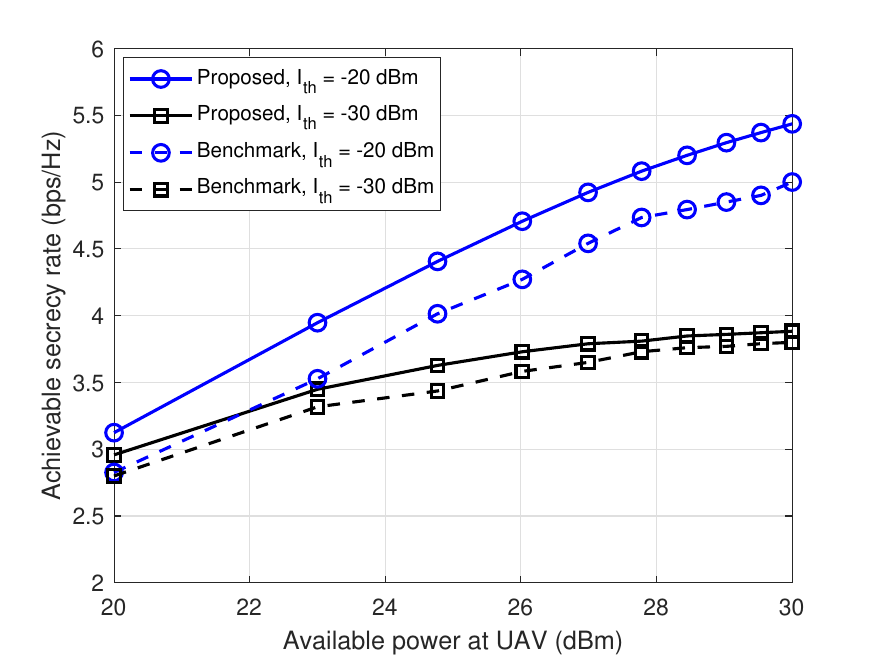}
\caption{Available power of UAV versus achievable secrecy rate of the system considering different interference temperatures, considering $M=32$.}
\label{Fig2}
\end{figure}

Fig. \ref{Fig2} shows the relationship between the achievable secrecy rate and the available transmit power at the UAV under different interference temperature limits to the primary network, i.e., \(I_{\text{th}} = -20\) dBm and \(I_{\text{th}} = -30\) dBm. The results indicate that increasing the UAV's transmit power generally enhances the secrecy rate of the proposed framework and benchmark method with both limits \(I_{\text{th}}\). However, when the interference limit is strict (\(I_{\text{th}} = -30\) dBm), the secrecy rate initially increases with UAV power but eventually saturates as the optimization reaches the interference temperature constraint, restricting further performance gains. In contrast, for a higher interference limit (\(I_{\text{th}} = -20\) dBm), the secrecy rate continues to improve with increasing UAV power, as the system has more flexibility in optimizing power allocation and beamforming. In all cases, the proposed framework performs better than the benchmark method. This highlights the trade-off between improving secrecy performance and maintaining acceptable interference levels to the primary network in cognitive radio-enabled NTNs.

\begin{figure}[!t]
\centering
\includegraphics [width=.48\textwidth]{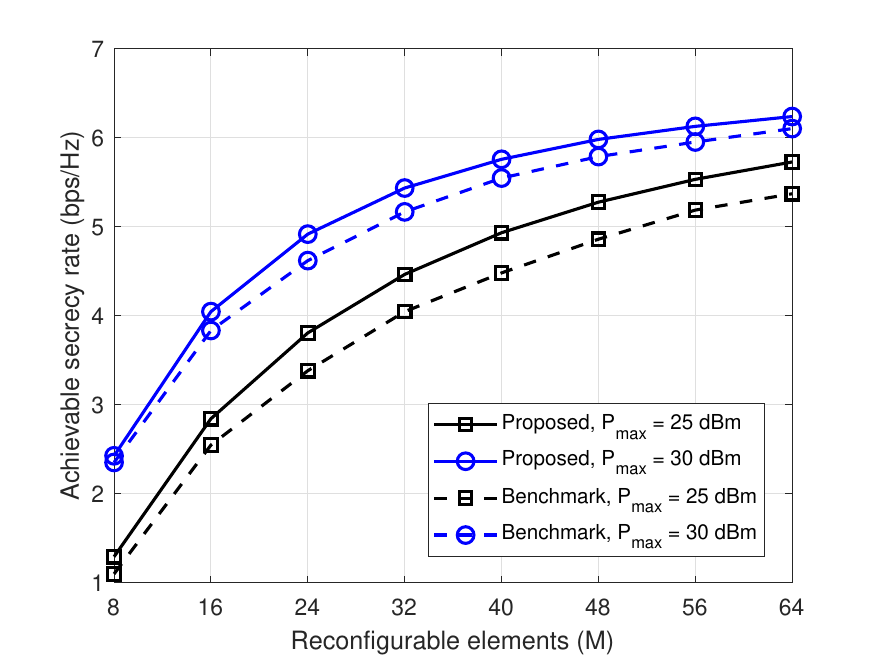}
\caption{Number of reconfigurable elements versus achievable secrecy rate of the system considering different transmit power of UAV, considering $I_{\text{th}}=10^{-5}$.}
\label{Fig3}
\end{figure}

Fig. \ref{Fig3} plots the achievable secrecy rate versus variability of BD-RIS elements under different transmit power levels of the UAV, i.e., \(P_{\text{max}} = 25\) dBm and \(P_{\text{max}} = 30\) dBm. It can be seen that increasing the number of BD-RIS elements improves the achievable secrecy rate of the secondary system of the proposed framework and benchmark method. With \(P_{\text{max}} = 30\) dBm, the secrecy rate initially increases with UAV power, but the improvement gain approaches saturation as the optimization reaches the interference temperature constraint, slowing down further performance gains. On the other hand, for a lower UAV power (\(P_{\text{max}} = 25\) dBm), the secrecy rate continues to improve with increasing BD-RIS elements, as the system has more flexibility in optimizing BD-RIS beamforming. Moreover, the proposed framework outperforms the benchmark method with a random phase shift of BD-RIS. This highlights the importance of optimizing both UAV transmit power and BD-RIS phase shift to maximize the system's secrecy performance in cognitive radio-enabled NTNs.

\section{Conclusion} \label{sec_5}
This work explored the physical layer security of a cognitive radio-enabled NTN, where a transmissive BD-RIS-mounted UAV serves as a secondary transmitter alongside a HAPS-based primary network. To maximize the secrecy rate while adhering to interference constraints, a joint optimization framework that efficiently balances power allocation and BD-RIS phase shift optimization is proposed. Given the inherent non-convexity of the problem, a novel alternating optimization framework, leveraging a switch-based low-complexity power allocation strategy and Stiefel manifold optimization for phase adjustment, is presented. Simulation results validate the superiority of the proposed scheme, demonstrating significant secrecy rate improvements over the conventional diagonal RIS design, highlighting the potential of BD-RIS for securing multi-layered NTN ecosystem.

\section*{Appendix A: Detailed Partial Derivatives}
Let us recall Equation (\ref{new}) as:
\begin{align} L(\boldsymbol{\Phi}, \lambda) = R_s^{\text{sec}} - \lambda \left( |\mathbf{g} \mathbf{\Phi}|^2 P_s - I_{\text{th}} \right),\label{31}  \end{align}
where
\begin{align}
R_s^{\text{sec}} =[\log_2 \left( 1 + \gamma_s \right) - \log_2 \left( 1 + \gamma_e \right)]^+
\end{align}
and
\begin{align}
\gamma_s = \frac{|\mathbf{h}_s \mathbf{\Phi}|^2 P_s}{\sigma_s^2 + |f_s|^2 Q_p}, \quad \gamma_e = \frac{|\mathbf{h}_e \mathbf{\Phi}|^2 P_s}{\sigma_e^2 + |f_e|^2 Q_p}.
\end{align}
The partial derivative of (\ref{31}) with respect to \(\mathbf{\Phi}\) can be expressed as:
\begin{align}
\frac{\partial L(\boldsymbol{\Phi}, \lambda)}{\partial \mathbf{\Phi}}& = \frac{\partial}{\partial \mathbf{\Phi}} \left[ \log_2 \left( 1 + \gamma_s \right) - \log_2 \left( 1 + \gamma_e \right) \right]\nonumber\\ & -\lambda\frac{\partial}{\partial \Phi}(|\mathbf{g} \mathbf{\Phi}|^2 P_s).
\end{align}
Using the chain rule, the derivative of the objective function can be written as:
\begin{align}
\frac{\partial R^{\text{sec}}_s}{\partial \mathbf{\Phi}} = \frac{1}{\ln(2)} \left( \Gamma_s \frac{\partial \gamma_s}{\partial \mathbf{\Phi}} - \Gamma_e \frac{\partial \gamma_e}{\partial \mathbf{\Phi}} \right).
\end{align}
where $\Gamma_s=\frac{1}{1 + \gamma_s}$ and $\Gamma_e=\frac{1}{1 + \gamma_e}$.
The derivative of \(\gamma_s\) with respect to \(\mathbf{\Phi}\) can be stated as:
\begin{align}
\frac{\partial \gamma_s}{\partial \mathbf{\Phi}} = \frac{P_s}{\sigma_s^2 + |f_s|^2 Q_p} \times \frac{\partial |A_s|^2}{\partial \mathbf{\Phi}}.
\end{align}
where \(A_s = \mathbf{h}_s \mathbf{\Phi}\). Since \(|A_s|^2 = A_s A_s^H\), we have:
\begin{align}
   \frac{\partial |A_s|^2}{\partial \mathbf{\Phi}} = 2 \mathbf{h}_s^H A_s.
\end{align}
   Therefore
\begin{align}
   \frac{\partial \gamma_s}{\partial \mathbf{\Phi}} = \frac{2 P_s \mathbf{h}_s^H A_s}{\sigma_s^2 + |f_s|^2 Q_p}.
\end{align}
Next, the derivative of \(\gamma_e\) with respect to \(\mathbf{\Phi}\) is computed as:
 \begin{align}
   \frac{\partial \gamma_e}{\partial \mathbf{\Phi}} = \frac{P_s}{\sigma_e^2 + |f_e|^2 Q_p} \times \frac{\partial |A_e|^2}{\partial \mathbf{\Phi}}.
\end{align}
where \(A_e = \mathbf{h}_e \mathbf{\Phi}\). Since \(|A_e|^2 = A_e A_e^H\), we have:
\begin{align}
\frac{\partial |A_e|^2}{\partial \mathbf{\Phi}} = 2 \mathbf{h}_e^H A_e.
 \end{align}
Therefore
\begin{align}
   \frac{\partial \gamma_e}{\partial \mathbf{\Phi}} = \frac{2 P_s \mathbf{h}_e^H A_e}{\sigma_e^2 + |f_e|^2 Q_p}.
\end{align}
Substituting \(\frac{\partial \gamma_s}{\partial \mathbf{\Phi}}\) and \(\frac{\partial \gamma_e}{\partial \mathbf{\Phi}}\) into the expression for \(\frac{\partial R_s^{\text{sec}}}{\partial \mathbf{\Phi}}\), it can be written as:
\begin{align}
\frac{\partial R^{\text{sec}}_s}{\partial \mathbf{\Phi}}=\frac{1}{\ln(2)} \left( \Gamma_s \frac{2 P_s \mathbf{h}_s^H A_s}{\sigma_s^2 + |f_s|^2 Q_p} - \Gamma_e \frac{2 P_s \mathbf{h}_e^H A_e}{\sigma_e^2 + |f_e|^2 Q_p} \right).
\end{align}
Simplifying the expression, we can express it as:
\begin{align}
\frac{\partial R^{\text{sec}}_s}{\partial \mathbf{\Phi}}&=\frac{2 P_s}{\ln(2)} \bigg( \frac{\mathbf{h}_s^H A_s}{(1 + \gamma_s)(\sigma_s^2 + |f_s|^2 Q_p)} \nonumber\\
& - \frac{\mathbf{h}_e^H A_e}{(1 + \gamma_e)(\sigma_e^2 + |f_e|^2 Q_p)} \bigg).
\end{align}
Accordingly, we calculate the derivative of the second term in the Lagrangian function as:
\begin{align}
-\lambda\frac{\partial}{\partial \Phi}(|\mathbf{g} \mathbf{\Phi}|^2 P_s).
\end{align}
We expand the square magnitude, and then compute the derivation such as:
\begin{align}
-\lambda P_s \frac{\partial}{\partial \Phi}((\mathbf{g} \mathbf{\Phi})(\mathbf{g} \mathbf{\Phi})^H).
\end{align}
The derivation is given as:
\begin{align}
-2\lambda P_s (\mathbf{g}^H \mathbf{\Phi})\mathbf{g}.
\end{align}
Combining this with the derivation of the objective function, it can be expressed as:
\begin{align}
&\frac{\partial L(\Phi,\lambda)}{\partial \mathbf{\Phi}}=\frac{2 P_s}{\ln(2)} \bigg( \frac{\mathbf{h}_s^H A_s}{(1 + \gamma_s)(\sigma_s^2 + |f_s|^2 Q_p)} \nonumber\\
& - \frac{\mathbf{h}_e^H A_e}{(1 + \gamma_e)(\sigma_e^2 + |f_e|^2 Q_p)} \bigg)-2\lambda P_s (\mathbf{g}^H \mathbf{\Phi})\mathbf{g}.
\end{align}

\bibliographystyle{IEEEtran}
\bibliography{Wali_EE}

\end{document}